\documentclass[12pt,a4paper]{article}

\usepackage{latexsym,amssymb,amsmath,amsbsy,epsfig}

\newcommand{\be}{\begin{equation}}
\newcommand{\ee}{\end{equation}}
\newcommand{\ba}{\begin{eqnarray}}
\newcommand{\ea}{\end{eqnarray}}
\newcommand{\bs}{\begin{subequations}}
\newcommand{\es}{\end{subequations}}
\newcommand{\no}{\nonumber\\}

\newcommand{\diag}{\mbox{diag}}
\newcommand{\Z}{\mathbb{Z}}

\begin{document}
\renewcommand{\thefootnote}{\fnsymbol{footnote}}

\title{
\normalsize \hfill CFTP/13-009
\\
\normalsize \hfill UWThPh-2013-9
\\*[7mm]
\LARGE Five models for lepton mixing}

\author{
P.M.\ Ferreira,$^{(1,2)}$\thanks{E-mail: ferreira@cii.fc.ul.pt} \
L.\ Lavoura,$^{(3)}$\thanks{E-mail: balio@cftp.ist.utl.pt} \
and P.O.\ Ludl$ \, ^{(4)}$\thanks{E-mail: patrick.ludl@univie.ac.at}
\\[3mm]
$^{(1)} \! $
\small Instituto Superior de Engenharia de Lisboa \\
\small 1959-007 Lisbon, Portugal
\\[1mm]
$^{(2)} \! $
\small Centre for Theoretical and Computational Physics,
University of Lisbon \\
\small 1649-003 Lisbon, Portugal
\\[1mm]
$^{(3)} \! $
\small Technical University of Lisbon, Instituto Superior T\'ecnico, CFTP \\
\small 1049-001 Lisbon, Portugal
\\[1mm]
$^{(4)} \! $
\small University of Vienna, Faculty of Physics \\
\small Boltzmanngasse 5, A--1090 Vienna, Austria
\\[3mm]
}

\date{5 April 2013}

\maketitle

\begin{abstract}
We produce five flavour models for the lepton sector.
All five models fit perfectly well---at the $1 \sigma$ level---the
existing data on the neutrino mass-squared differences
and on the lepton mixing angles.
The models are based on the type~I seesaw mechanism,
on a $\Z_2$ symmetry for each lepton flavour,
and either on a
(spontaneously broken)
symmetry under the interchange of two lepton flavours or on a
(spontaneously broken)
$CP$ symmetry incorporating that interchange---or on both symmetries
simultaneously.
Each model makes definite predictions both for the scale of the neutrino masses
and for the phase $\delta$ in lepton mixing; the fifth model also predicts
a correlation between the lepton mixing angles $\theta_{12}$ and $\theta_{23}$.
\end{abstract}

\newpage

\renewcommand{\thefootnote}{\arabic{footnote}}

\section{Introduction}

The problem of explaining the observed features of lepton mixing
became significantly more awkward with the recent definitive
(more than $5\sigma$)
indications for a non-zero mixing angle $\theta_{13}$~\cite{daya-reno}.
These indications have rendered obsolete
the idea of a $\mu$--$\tau$ interchange symmetry
in the neutrino mass matrix~\cite{original}.
Moreover,
a not so well-established
($2 \sigma$)
indication that the mixing angle $\theta_{23}$
is not maximal\footnote{In this paper we use exclusively
the global fit of the neutrino-oscillation data
by Fogli \textit{et al.}~\cite{fogli}.
Other fits are contained in refs.~\cite{tortola,schwetz}.}
suggests that the idea of a $CP$ symmetry
incorporating the $\mu$--$\tau$ interchange~\cite{cp}
should be discarded too.

At this juncture,
one may either search for other ideas
or try and somehow modify the approaches
mentioned in the previous paragraph
in order to make them compatible with the data.
In this paper we take the latter path.
Specifically,
while in the models of refs.~\cite{original,discrete}
the neutrino Dirac mass matrix $M_D$
was of the form $\diag \left( a, b, b \right)$
and in the model of ref.~\cite{cp} it was $\diag \left( a, b, b^\ast \right)$,
in this paper we present models where $M_D = \diag \left( a, b, c \right)$.
We shall see that this still allows us to retain
some predictive power.
In order to enhance that power,
we also apply in this paper an idea 
originally proposed in ref.~\cite{discrete},
which allows us to obtain one
vanishing
off-diagonal matrix element
in the
inverse of the
light-neutrino Majorana mass matrix $M$
(in the basis where the charged-lepton mass matrix is diagonal).

In general,
the matrix $M$,
which is symmetric,
contains nine
physical parameters:
the six moduli of its matrix elements and three rephasing-invariant phases.
Those nine parameters correspond to nine observables:
three neutrino masses,
three mixing angles,
one Dirac phase ($\delta$) in the mixing,
and two Majorana phases among the three neutrino masses.
The first four models that we present in this paper
have only five physical parameters and are thus comparable,
in predictive power,
to models or \textit{Ans\"atze}\/
with two `texture zeros'~\cite{marfatia,lavoura}.
Those models are able to predict
$\delta$ and the absolute scale of the
neutrino masses.\footnote{We omit predictions for the two Majorana phases,
since they are,
in general,
experimentally irrelevant.}
Moreover,
our models provide excellent fits
to the existing data on the three mixing angles
and on the two neutrino mass-squared differences.
Our fifth model has only four physical parameters and is
able to predict one correlation between two mixing angles. 

All the models that we propose in this paper
utilize the type~I seesaw mechanism
with three right-handed neutrinos.
Thus,
the leptonic multiplets in our models are the following:
\begin{itemize}
\item Three right-handed charged-lepton gauge-$SU(2)$ singlets $\alpha_R$,
where $\alpha$ may take the values $e$,
$\mu$,
and $\tau$;
\item three right-handed neutrino singlets $\nu_{\alpha R}$;
\item three left-handed doublets
$D_{\alpha L} = \left( \nu_{\alpha L}, \ \alpha_L \right)^T$.
\end{itemize}
Our models also have a very simple scalar sector:
\begin{itemize}
\item Two Higgs doublets
$\phi_{1,2} = \left( \phi_{1,2}^+, \ \phi_{1,2}^0 \right)^T$.
Their conjugate doublets
are $\tilde \phi_{1,2} = \left( {\phi_{1,2}^0}^\ast, \
- \phi_{1,2}^- \right)^T$;
\item two {\em real}\/ gauge singlets,
$\chi_1$ and $\chi_2$.
\end{itemize}

The outline of this paper is as follows.
In section~\ref{matrix} we introduce a matrix $A$,
derived from the matrix $M$
but which does not suffer from the rephasing ambiguity of the latter;
it turns out that most predictions of our models
can be stated in terms of $A$-matrix elements.
In section~\ref{sec:first} we present our first model,
which uses a $\mu$--$\tau$ interchange symmetry.
In section~\ref{sec:second} we present our second model,
which uses a $CP$ symmetry incorporating the $\mu$--$\tau$ interchange.
In section~\ref{sec:third} we note that in both previous models
one may use an $e$--$\tau$ interchange
instead of the $\mu$--$\tau$ interchange
and still obtain a perfect fit to the data;
we moreover note that,
by combining the $e$--$\tau$ interchange symmetry
with the corresponding generalized $CP$ symmetry,
one obtains a fifth model that is still compatible with the data.
In section~6 we elaborate, by means of a fit to the phenomenological
data and of scatter plots,
the predictions of our models.
We briefly summarize our achievements in section~\ref{sec:conclusions}.

\section{The matrix $A$}
\label{matrix}

Let
\be
\label{M}
M = \left( \begin{array}{ccc}
a & f & e \\ f & b & d \\ e & d & c
\end{array} \right)
\ee
be a symmetric $3\times 3$ matrix.
We assume $M$ to be invertible.
Then,
\be
M^{-1} = \frac{1}{\det{M}} \left( \begin{array}{ccc}
b c - d^2 & d e - c f & d f - b e \\
d e - c f & a c - e^2 & e f - a d \\
d f - b e & e f - a d & a b - f^2
\end{array} \right).
\ee
We define the matrix $A$,
which is a symmetric $3\times 3$ matrix too,
through
\be
A_{\alpha \beta} = M_{\alpha \beta} \left( M^{-1} \right)_{\alpha \beta},
\ee
where we do not imply summation
over $\alpha$ and $\beta$.\footnote{In this section,
Greek-letter indices have range $\left\{ 1, 2, 3 \right\}$.
In the other sections of this paper,
$M$ is interpreted as the neutrino mass matrix
and those indices indicate the lepton flavours
$e$, $\mu$ and $\tau$.}
Then,
\be
\label{A}
A = \frac{1}{\det{M}} \left( \begin{array}{ccc}
a b c - a d^2 & d e f - c f^2 & d e f - b e^2 \\
d e f - c f^2 & a b c - b e^2 & d e f - a d^2 \\
d e f - b e^2 & d e f - a d^2 & a b c - c f^2
\end{array} \right).
\ee
An advantage of the matrix $A$ over the matrix $M$ is that,
when $M$ gets rephased through
\be
\label{reph}
M_{\alpha \beta} \to e^{i \left( \psi_\alpha + \psi_\beta \right)} M_{\alpha \beta},
\ee
the matrix $A$ remains invariant.

Since
\be
\det{M} = a b c + 2 d e f - a d^2 - b e^2 - c f^2,
\ee
the matrix $A$ in eq.~(\ref{A}) satisfies
\be
\label{sum}
\sum_{\alpha = 1}^3 A_{\alpha \beta} = 1
\ee
for any value of $\beta$.
It follows from eq.~(\ref{sum}) that only three elements of $A$
are linearly independent.
Therefore $A$
has only
six parameters:
three moduli and three phases.\footnote{The original matrix $M$,
after allowing for the rephasing~(\ref{reph}),
has nine parameters:
six moduli and three phases.
So,
$A$ contains less information than $M$.}

Equation~(\ref{sum}) has several consequences.
Let $\alpha \neq \beta \neq \gamma \neq \alpha$.
Then,
\be
\label{first}
A_{\gamma \alpha} = A_{\gamma \beta}
\ \Leftrightarrow \
A_{\alpha \alpha} = A_{\beta \beta},
\ee
which holds because $A_{\alpha \alpha} = 1 - A_{\beta \alpha} - A_{\gamma \alpha}$,
$A_{\beta \beta} = 1 - A_{\alpha \beta} - A_{\gamma \beta}$,
and $A$ is symmetric.
Another consequence is
\be
A_{\gamma \alpha} = A_{\gamma \beta}^\ast
\ \Rightarrow \
A_{\gamma \gamma} = A_{\gamma \gamma}^\ast,
\ee
which holds because $A_{\gamma \gamma} = 1 - A_{\alpha \gamma} - A_{\beta \gamma}$.
Still another consequence of eq.~(\ref{sum}) is
\be
A_{\gamma \alpha} = A_{\gamma \beta}^\ast
\, \wedge \,
A_{\alpha \alpha} = A_{\beta \beta}^\ast
\ \Rightarrow \
A_{\alpha \beta} = A_{\alpha \beta}^\ast.
\ee
Finally,
one further consequence of eq.~(\ref{sum}) is
\be
A_{\gamma \alpha} = A_{\gamma \beta}^\ast
\, \wedge \,
A_{\alpha \beta} = A_{\alpha \beta}^\ast
\ \Rightarrow \
A_{\alpha \alpha} = A_{\beta \beta}^\ast.
\ee

\section{Model~1}
\label{sec:first}

Our first model has four $\Z_2$ symmetries.
The first three of them are
\begin{itemize}
\item $\Z_2^{(e)}: \ D_{eL},\ e_R,\ \nu_{eR},\ \chi_1,\ \chi_2$
change sign;
\item $\Z_2^{(\mu)}: \ D_{\mu L},\ \mu_R,\ \nu_{\mu R},\ \chi_1$
change sign;
\item $\Z_2^{(\tau)}: \ D_{\tau L},\ \tau_R,\ \nu_{\tau R},\ \chi_2$
change sign.
\end{itemize}
These three symmetries are broken spontaneously at the high (seesaw) scale
when $\chi_{1,2}$ acquire vacuum expectation values (VEVs)
\be
\left\langle 0 \left| \chi_1 \right| 0 \right\rangle \equiv U \cos{\vartheta},
\quad
\left\langle 0 \left| \chi_2 \right| 0 \right\rangle \equiv U \sin{\vartheta}.
\ee
The fourth $\Z_2$ symmetry of model~1 is
\be
\label{z2int}
\Z_2^\mathrm{(int)}:\ D_{\mu L} \leftrightarrow D_{\tau L},\
\mu_R \leftrightarrow \tau_R,\
\nu_{\mu R} \leftrightarrow \nu_{\tau R},\
\chi_{1} \leftrightarrow \chi_{2},\
\phi_2 \to - \phi_2.
\ee
This symmetry is broken spontaneously at the low
(Fermi or lower\footnote{Our Higgs doublets $\phi_{1,2}$
are not necessarily the ones which couple to the quarks.
Their VEVs may be much lower
(and their masses much higher)
than the Fermi scale,
suppressed for instance through a type~II seesaw mechanism~\cite{radovcic}.
Alternatively,
our model~1 may be viewed as an ordinary two-Higgs-doublet model
furnished with a $\Z_2$ symmetry under which $\phi_2 \to - \phi_2$;
that symmetry may apply to the quark sector
in a variety of ways~\cite{2HDM},
for instance by inverting the signs of the right-handed down-type quarks.
It has been shown~\cite{show} that these models are quite capable of describing
the phenomenology and experimental constraints that arise from the 
recent observations of a Higgs-like particle~\cite{lhc}.})
scale
when $\phi_2^0$ acquires a VEV.

The Lagrangian of Majorana mass terms is then of the form
\be \label{uziuy}
\mathcal{L}_\mathrm{Maj} =
- \frac{1}{2} \left[ m \bar \nu_{eR} C \bar \nu_{eR}^T
+ m' \left( \bar \nu_{\mu R} C \bar \nu_{\mu R}^T +
\bar \nu_{\tau R} C \bar \nu_{\tau R}^T \right) \right] + \mathrm{H.c.}
\ee
The Yukawa Lagrangian is
\ba
\mathcal{L}_\mathrm{Yuk} &=&
- y_1 \bar D_{eL} e_R \phi_1
\no & &
- y_2 \left( \bar D_{\mu L} \mu_R + \bar D_{\tau L} \tau_R \right) \phi_1
- y_3 \left( \bar D_{\mu L} \mu_R - \bar D_{\tau L} \tau_R \right) \phi_2
\no & &
- y_4 \bar D_{eL} \nu_{eR} \tilde \phi_1
\no & &
- y_5 \left( \bar D_{\mu L} \nu_{\mu R} + \bar D_{\tau L} \nu_{\tau R} \right)
\tilde \phi_1
- y_6 \left( \bar D_{\mu L} \nu_{\mu R} - \bar D_{\tau L} \nu_{\tau R} \right)
\tilde \phi_2
\no & &
- y_7 \left( \chi_1 \bar \nu_{\mu R} + \chi_2 \bar \nu_{\tau R} \right)
C \bar \nu_{eR}^T
+ \mathrm{H.c.}, \label{jbuit}
\ea
where $C$ is the charge-conjugation matrix in Dirac space.
So the charged-lepton mass matrix is diagonal,
\be
M_\ell = \mathrm{diag} \left(
y_1 v_1, y_2 v_1 + y_3 v_2, y_2 v_1 - y_3 v_2
\right),
\ee
where $v_k = \left\langle 0 \left| \phi_k^0 \right| 0 \right\rangle$
for $k = 1, 2$.
The 
neutrino Dirac mass matrix
is diagonal too:
\be
\label{abc}
M_D = \diag \left( a, b, c \right),
\ee
where $a = y_4^\ast v_1$,
$b =  y_5^\ast v_1 + y_6^\ast v_2$,
and $c = y_5^\ast v_1 - y_6^\ast v_2$.
The Majorana mass matrix of the right-handed neutrinos is given by
\be
M_R = \left( \begin{array}{ccc}
m & y_7 U \cos{\vartheta} & y_7 U \sin{\vartheta} \\
y_7 U \cos{\vartheta} & m'& 0 \\
y_7 U \sin{\vartheta} & 0 & m'
\end{array} \right).
\ee
The expression for the effective light-neutrino Majorana mass matrix
is\footnote{In section~\ref{matrix},
$M$ was a generic $3 \times 3$ symmetric,
non-singular matrix.
>From now on,
$M$ will specifically denote the light-neutrino Majorana mass matrix
in the basis where the charged-lepton mass matrix is diagonal.}
\be
\label{seesaw}
M = - M_D^T M_R^{-1} M_D.
\ee
Since $M_D$ is diagonal and $\left( M_R \right)_{\mu\tau} = 0$,
it immediately follows that
\be
\label{cond1}
\left( M^{-1} \right)_{\mu\tau} = 0.
\ee

Taking into account the imposed symmetries,
one finds the scalar potential
\ba
V &=& \sum_{k=1}^2 \left[ \mu_k \phi_k^\dagger \phi_k
+ \lambda_k \left( \phi_k^\dagger \phi_k \right)^2 \right]
\no & &
+ \lambda_3\, \phi_1^\dagger \phi_1\, \phi_2^\dagger \phi_2
+ \lambda_4\, \phi_1^\dagger \phi_2\, \phi_2^\dagger \phi_1
+ \left[ \lambda_5 \left( \phi_1^\dagger \phi_2 \right)^2
+ \mathrm{H.c.} \right]
\no & &
+ \left( \mu_\chi + \sum_{k=1}^2 \bar \lambda_k\, \phi_k^\dagger \phi_k \right)
\left( \chi_1^2 + \chi_2^2 \right)
+ \lambda_\chi \left( \chi_1^2 + \chi_2^2 \right)^2
\no & &
+ \lambda_\chi' \chi_1^2 \chi_2^2
+ \left(
\lambda
\phi_1^\dagger \phi_2 + \mathrm{H.c.} \right)
\left( \chi_1^2 - \chi_2^2 \right). \label{ivuty}
\ea
We assume that $U$ is at the high scale,
while $v = \left( \left| v_1 \right|^2 + \left| v_2 \right|^2 \right)^{1/2}$
lies at the low scale.\footnote{This of course requires
some finetuning in $V$.
Namely,
$\bar \lambda_{1,2}$ must be very small,
of order $v^2/U^2$,
in order that $\mu_1$ and $\mu_2$ do not receive corrections
of seesaw magnitude when $\chi_{1,2}$ acquire VEVs.
This is an ordinary `hierarchy problem',
to which we cannot offer a solution.}
The vacuum potential for $U$ and $\vartheta$ is
\be
\label{v0}
V_0 =
\left( \mu_\chi + \sum_{k=1}^2 \bar \lambda_k \left| v_k \right|^2 \right) U^2
+ \lambda_\chi U^4
+ \frac{\lambda'_\chi}{4}\, U^4 \sin^2{2 \vartheta}
+ 2 \Re \left(
\lambda
v_1^\ast v_2 \right) U^2 \cos{2 \vartheta}.
\ee
Unless
$\lambda$
is very large,
of order $U^2/v^2$,
the last term in the right-hand side of eq.~(\ref{v0})
is much smaller than the term just before it
and may be neglected.\footnote{This is the same approximation
as in eq.~(\ref{seesaw}),
which neglects terms of order $v^2 / U^2$~\cite{GLseesaw} too.}
Then $\vartheta$ will be either $n \pi/2$ or $(2n+1)\pi/4$
(with integer $n$)
depending on whether $\lambda'_\chi$ is,
respectively,
positive or negative.
We assume that $\lambda'_\chi$ is negative and that $\vartheta = \pi/4$,
\textit{i.e.}\ that
$\left\langle 0 \left| \chi_1 \right| 0 \right\rangle
= \left\langle 0 \left| \chi_2 \right| 0 \right\rangle$.
This equality holds to $\mathcal{O} \left( v^2/U^2 \right)$
when
$\lambda$ and $\lambda^\prime_\chi$ are of the same order.

Thus,
\be
\label{eme}
M^{-1} = -\,
\diag \left( \frac{1}{a},\, \frac{1}{b},\, \frac{1}{c} \right)
\left( \begin{array}{ccc}
m & m^{\prime\prime} & m^{\prime\prime} \\
m^{\prime\prime} & m' &
0 \\ m^{\prime\prime} & 0 & m'
\end{array} \right)
\diag \left( \frac{1}{a},\, \frac{1}{b},\, \frac{1}{c} \right),
\ee
where $m^{\prime\prime} = y_7 U \left/ \sqrt{2} \right.$.
Hence,
\ba
\label{eme-1}
M &=&
- \frac{1}{m^\prime \left(m m^\prime - 2 {m^{\prime\prime}}^2 \right)}\
\diag \left( a, b, c \right)
\no & & \times \left( \begin{array}{ccc}
{m^\prime}^2 & - m^\prime m^{\prime\prime} &  - m^\prime m^{\prime\prime} \\
- m^\prime m^{\prime\prime} & m m'- {m^{\prime\prime}}^2 & {m^{\prime\prime}}^2 \\
- m^\prime m^{\prime\prime} & {m^{\prime\prime}}^2 & m m'- {m^{\prime\prime}}^2
\end{array} \right)
\diag \left( a, b, c \right).
\hspace*{5mm}
\ea
It follows from eqs.~(\ref{eme}) and~(\ref{eme-1})
that\footnote{$A_{\mu\mu} = A_{\tau\tau}$
also follows from eqs.~(\ref{eme}) and~(\ref{eme-1})
but is equivalent to eq.~(\ref{cond2}),
\textit{cf.}\ eq.~(\ref{first}).}
\ba
\label{cond2}
A_{e\mu} = A_{e\tau}.
\ea
We conclude that model~1 leads to two complex conditions on $M$,
eqs.~(\ref{cond1}) and~(\ref{cond2}).
Therefore,
the matrix $M$ in this model contains five physical parameters,
as opposed to nine physical parameters in the general case.

\section{Model~2}
\label{sec:second}

Our second model also has the symmetries $\Z_2^{(e)}$,
$\Z_2^{(\mu)}$,
and $\Z_2^{(\tau)}$.
However,
now $\Z_2^{(\mathrm{int})}$ is substituted by the $CP$ symmetry
\be
\label{cpcp}
\begin{array}{rcl}
& &
\left\{
\begin{array}{rcl}
D_{eL} (x) &\to& \gamma_0 C \bar D_{eL}^T (\bar x)
\\*[1mm]
D_{\mu L} (x) &\to& \gamma_0 C \bar D_{\tau L}^T (\bar x)
\\*[1mm]
D_{\tau L} (x) &\to& \gamma_0 C \bar D_{\mu L}^T (\bar x)
\end{array}
\right.,
\quad
\left\{
\begin{array}{rcl}
e_R (x) &\to& \gamma_0 C \bar e_R^T (\bar x)
\\*[1mm]
\mu_R (x) &\to& \gamma_0 C \bar \tau_R^T (\bar x)
\\*[1mm]
\tau_R (x) &\to& \gamma_0 C \bar \mu_R^T (\bar x)
\end{array}
\right.,
\\*[-2mm]
CP: & &
\\*[-2mm]
& &
\left\{
\begin{array}{rcl}
\nu_{eR} (x) &\to& \gamma_0 C \bar \nu_{eR}^T (\bar x)
\\*[1mm]
\nu_{\mu R} (x) &\to& \gamma_0 C \bar \nu_{\tau R}^T (\bar x)
\\*[1mm]
\nu_{\tau R} (x) &\to& \gamma_0 C \bar \nu_{\mu R}^T (\bar x)
\end{array}
\right.,
\quad
\left\{
\begin{array}{rcl}
\chi_1 (x) &\to& \chi_2 (\bar x)
\\*[1mm]
\chi_2 (x) &\to& \chi_1 (\bar x)
\\*[1mm]
\phi_1 (x) &\to& \phi_1^\ast (\bar x)
\\*[1mm]
\phi_2 (x) &\to& - \phi_2^\ast (\bar x)
\end{array}
\right.,
\end{array}
\ee
where $x = \left( t, \vec r \right)$ and $\bar x = \left( t, - \vec r \right)$.
Notice that,
since $\chi_1$ and $\chi_2$ are gauge-singlet fields,
they have no electroweak interactions,
hence $CP$ may be arbitrarily defined for them;
we have opted for
$\chi_1\,$\raisebox{-2pt}{$\stackrel{CP}{\leftrightarrow}\,$}$\chi_2$.
Also notice that the Higgs doublets $\phi_1$ and $\phi_2$
transform with opposite signs under $CP$.

We now have
\be \label{cjuit}
\mathcal{L}_\mathrm{Maj} =
- \frac{1}{2} \left[ m \bar \nu_{eR} C \bar \nu_{eR}^T
+ m' \bar \nu_{\mu R} C \bar \nu_{\mu R}^T
+ {m'}^\ast \bar \nu_{\tau R} C \bar \nu_{\tau R}^T
\right] + \mathrm{H.c.},
\ee
where $m$ is real,
and
\ba
\mathcal{L}_\mathrm{Yuk} &=&
- y_1 \bar D_{eL} e_R \phi_1
\no & &
- \left( y_2 \bar D_{\mu L} \mu_R + y_2^\ast \bar D_{\tau L} \tau_R \right) \phi_1
- \left( y_3 \bar D_{\mu L} \mu_R - y_3^\ast \bar D_{\tau L} \tau_R \right) \phi_2
\no & &
- y_4 \bar D_{eL} \nu_{eR} \tilde \phi_1
\no & &
- \left( y_5 \bar D_{\mu L} \nu_{\mu R} + y_5^\ast \bar D_{\tau L} \nu_{\tau R}
\right) \tilde \phi_1
- \left( y_6 \bar D_{\mu L} \nu_{\mu R} - y_6^\ast \bar D_{\tau L} \nu_{\tau R}
\right) \tilde \phi_2
\no & &
- \left( y_7 \chi_1\bar \nu_{\mu R}
+ y_7^\ast \chi_2 \bar \nu_{\tau R} \right) C \bar \nu_{eR}^T
+ \mathrm{H.c.}, \label{sviot}
\ea
where $y_1$ and $y_4$ are real.
So the charged-lepton mass matrix is once again diagonal,
\be
M_\ell = \diag \left( y_1 v_1, y_2 v_1 + y_3 v_2, y_2^\ast v_1 - y_3^\ast v_2
\right),
\ee
just as the neutrino Dirac mass matrix,
which is as in eq.~(\ref{abc}) with $a = y_4^\ast v_1$,
$b =  y_5^\ast v_1 + y_6^\ast v_2$,
and $c = y_5 v_1 - y_6 v_2$.
The Majorana mass matrix of the right-handed neutrinos is given by
\be
M_R = \left( \begin{array}{ccc}
m & y_7 U \cos{\vartheta} & y_7^\ast U \sin{\vartheta} \\
y_7 U \cos{\vartheta} & m'& 0 \\
y_7^\ast U \sin{\vartheta} & 0 & {m'}^\ast
\end{array} \right)
\ee
and eq.~(\ref{cond1}) still holds.

The scalar potential is
\ba
V &=& \sum_{k=1}^2 \left[ \mu_k \phi_k^\dagger \phi_k
+ \lambda_k \left( \phi_k^\dagger \phi_k \right)^2 \right]
\no & &
+ \lambda_3\, \phi_1^\dagger \phi_1\, \phi_2^\dagger \phi_2
+ \lambda_4\, \phi_1^\dagger \phi_2\, \phi_2^\dagger \phi_1
+\lambda_5 \left[ \left( \phi_1^\dagger \phi_2 \right)^2
+ \mathrm{H.c.} \right]
\no & &
+ \left( \mu_\chi + \sum_{k=1}^2 \bar \lambda_k\, \phi_k^\dagger \phi_k \right)
\left( \chi_1^2 + \chi_2^2 \right)
+ \lambda_\chi \left( \chi_1^2 + \chi_2^2 \right)^2
\no & &
+ \lambda_\chi' \chi_1^2 \chi_2^2
+ \left[
\lambda
\left( \phi_1^\dagger \phi_2 \chi_1^2
- \phi_2^\dagger \phi_1 \chi_2^2 \right)
+ \mathrm{H.c.} \right], \label{ivuro}
\ea
where $\lambda_5$ is real but
$\lambda$
is in general complex.
Once again,
if $\lambda_\chi'$ is negative then $\vartheta = \pi / 4$.
Thus,
\be
\label{neweme}
M^{-1} = -\,
\diag \left( \frac{1}{a},\, \frac{1}{b},\, \frac{1}{c} \right)
\left( \begin{array}{ccc}
m & m^{\prime\prime} & {m^{\prime\prime}}^\ast \\
m^{\prime\prime} & m' &
0 \\ {m^{\prime\prime}}^\ast & 0 & {m'}^\ast
\end{array} \right)
\diag \left( \frac{1}{a},\, \frac{1}{b},\, \frac{1}{c} \right),
\ee
where $m^{\prime\prime} = y_7 U \left/ \sqrt{2} \right.$
Hence,
\ba
\label{neweme-1}
M^{-1} &=& - \frac{1}{m \left| m^\prime \right|^2
- 2 \Re \left( {m^\prime}^\ast {m^{\prime \prime}}^2 \right)}\
\diag \left( a, b, c \right)
\no & & \times \left( \begin{array}{ccc}
\left| m^\prime\right|^2 & - {m^\prime}^\ast m^{\prime\prime} &
- m^\prime {m^{\prime\prime}}^\ast \\
- {m^\prime}^\ast m^{\prime\prime} & m {m'}^\ast- {{m^{\prime\prime}}^\ast}^2 &
\left| m^{\prime\prime} \right|^2 \\
- m^\prime {m^{\prime\prime}}^\ast & \left| m^{\prime\prime} \right|^2 &
m m'- {m^{\prime\prime}}^2
\end{array} \right)
\diag \left( a, b, c \right).
\no & &
\ea
Therefore this model predicts\footnote{It also predicts
$A_{\mu\mu} = A_{\tau\tau}^\ast$,
$A_{ee} = A_{ee}^\ast$,
and $A_{\mu\tau} = A_{\mu\tau}^\ast$,
but these three relations can be derived from eqs.~(\ref{cond1})
and~(\ref{complex}),
as was pointed out at the end of section~\ref{matrix}.}
\be
\label{complex}
A_{e\mu} = A_{e\tau}^\ast.
\ee
So this model has once again five parameters,
since it predicts the two complex eqs.~(\ref{cond1}) and~(\ref{complex}).

\section{Models~3, 4, and~5}
\label{sec:third}

In the models of the previous two sections
we have made the following assignment for the lepton flavours:
we have assumed that the lepton flavours being interchanged
either by the symmetry $\Z_2^\mathrm{(int)}$ of eq.~(\ref{z2int})
or by the $CP$ symmetry of eq.~(\ref{cpcp}) are the $\mu$ and $\tau$ flavours.
Two other choices are possible---the lepton flavours being interchanged
might be either $e$ and $\mu$ or $e$ and $\tau$.
We have tested those two other choices
against the experimental data and have found that
the second choice fits those data just as well as models~1 and~2 above.
Thus,
we define model~3 as predicting
\bs
\ba
\left( M^{-1} \right)_{e\tau} &=& 0,
\\
A_{e\mu} &=& A_{\mu\tau},
\ea
\es
and model~4 as predicting
\bs
\ba
\left( M^{-1} \right)_{e\tau} &=& 0,
\\
A_{e\mu} &=& A_{\mu\tau}^\ast.
\ea
\es

It is also possible to
\emph{simultaneously} impose invariance
under
$\Z_2^\mathrm{(int)}$ of eq.~(\ref{z2int})
and the
$CP$ symmetry of eq.~(\ref{cpcp}).
In that case,
\begin{itemize}
\item Equation~(\ref{uziuy}) holds with $m$ and $m^\prime$ real,
\textit{cf.}\ eq.~(\ref{cjuit}).
\item Equation~(\ref{jbuit}) holds with $y_{1\mbox{--}7}$ real,
\textit{cf.}\ eq.~(\ref{sviot}).
\item Equation~(\ref{ivuty}) holds with $\lambda_5$ and $\lambda$  real,
\textit{cf.}\ eq.~(\ref{ivuro}).
\end{itemize}
So,
finally one ends up with eq.~(\ref{eme}) with $m$,
$m^\prime$,
and $m^{\prime\prime}$ real.
The $CP$ symmetry 
may be broken spontaneously through VEVs $v_{1,2}$
with a relative complex phase and this makes $a$,
$b$,
and $c$ complex;
but that has no
relevance
for our predictions,
since $a, b, c$ drop out in the matrix $A$.

It turns out that the model delineated above
is unable to correctly fit the present data.
But the model in which the $e$ and $\mu$ flavours
are interchanged relative to the above works fine;
hence we define model~5 as predicting
\bs
\ba
\left( M^{-1} \right)_{e\tau} &=& 0,
\\
A_{e\mu} &=& A_{\mu\tau},
\\
A_{\mu\tau} &=& A_{\mu\tau}^\ast.
\ea
\es

\section{Observable predictions}

We have fitted models~1--5 of the previous sections
to the phenomenological data of ref.~\cite{fogli}.
It turns out that our models fit the data so perfectly
that we were able to
use just the $1\sigma$ intervals given in that paper.
Thus,
we have required that
\bs
\label{normal}
\ba
\label{dm}
m_2^2 - m_1^2 &\in& \left[ 7.32,\ 7.80 \right]
\times 10^{-5}\ \mathrm{eV}^2,
\\
\label{12}
\sin^2{\theta_{12}} &\in& \left[ 0.291,\ 0.325 \right],
\\
\label{Dm}
m_3^2 - \frac{m_1^2 + m_2^2}{2} &\in& \left[ 2.33,\ 2.49 \right]
\times 10^{-3}\ \mathrm{eV}^2,
\\
\label{13}
\sin^2{\theta_{13}} &\in& \left[ 0.0216,\ 0.0266 \right],
\\
\label{23}
\sin^2{\theta_{23}} &\in& \left[ 0.365,\ 0.410 \right],
\ea
\es
in the case of a
normal
neutrino mass spectrum.
If the neutrino mass spectrum is
inverted,
then the requirements~(\ref{dm}) and (\ref{12}) remain,
but the requirements~(\ref{Dm})--(\ref{23}) are substituted by
\bs
\label{inverted}
\ba
\label{Dminv}
\frac{m_1^2 + m_2^2}{2} - m_3^2 &\in& \left[ 2.31,\ 2.49 \right]
\times 10^{-3}\ \mathrm{eV}^2,
\\
\label{13inv}
\sin^2{\theta_{13}} &\in& \left[ 0.0219,\ 0.0267 \right],
\\
\label{23inv}
\sin^2{\theta_{23}} &\in& \left[ 0.370,\ 0.431 \right],
\ea
\es
respectively.
In eqs.~(\ref{normal}) and~(\ref{inverted}),
$m_{1,2,3}$ are the neutrino masses
and $\theta_{12,13,23}$ are the lepton mixing angles
as defined in the standard parameterization
in eq.~(13.79) of ref.~\cite{pdg}.
Notice that we have \emph{not}\/ used in our fits
the constraints in ref.~\cite{fogli} for the Dirac phase $\delta$
of lepton mixing,
which we chose instead to be an observable to be predicted by our models.
For comparison,
at the $1\sigma$ level the authors of ref.~\cite{fogli} found
\be\label{deltarange}
\begin{split}
& \delta \in [0.77\pi,\ 1.36\pi]
\quad \Rightarrow \quad
\mathrm{cos}\,\delta\in[-1,-0.43],
\\
& \delta \in [0.83\pi,\ 1.47\pi]
\quad \Rightarrow \quad
\mathrm{cos}\,\delta\in
[-1,-0.09]
\end{split}
\ee
for a normal and
for an
inverted spectrum, respectively.

We have found that all our models are able to fit the data
either in eqs.~(\ref{normal})---for a normal mass spectrum---or
in eqs.~(\ref{inverted})---for an inverted mass spectrum---or in
both of them---perfectly.
In models~1 to~4, the phase space for the five observables appears,
inside the intervals quoted in those equations,
uniformly filled in all our scatter plots,
so that no prediction for any of those observables looks warranted.
On the other hand,
predictions for the overall neutrino mass scale
and for the phase $\delta$ are possible.
Another quantity that one may predict
is the effective mass $m_{\beta\beta}\equiv |M_{ee}|$,
which is relevant for neutrinoless double beta decay.
In model~5,
moreover,
the angles $\theta_{12}$ and $\theta_{23}$ are correlated
and significantly restricted
relative to the intervals in eqs.~(\ref{12}) and~(\ref{23inv}),
respectively.

 \begin{figure}[ht]
\centerline{\epsfysize=6cm
\epsfbox{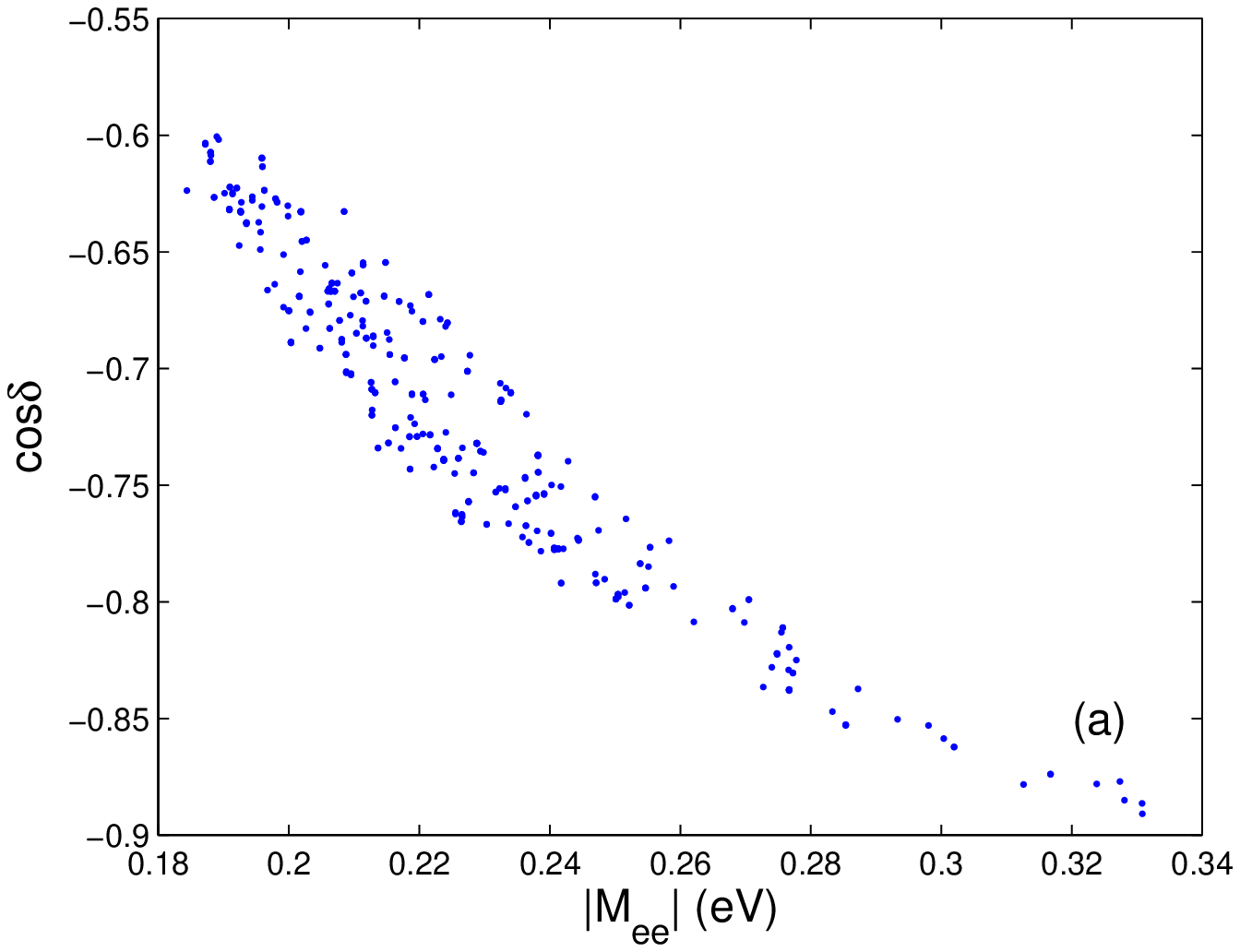}\epsfysize=6cm \epsfbox{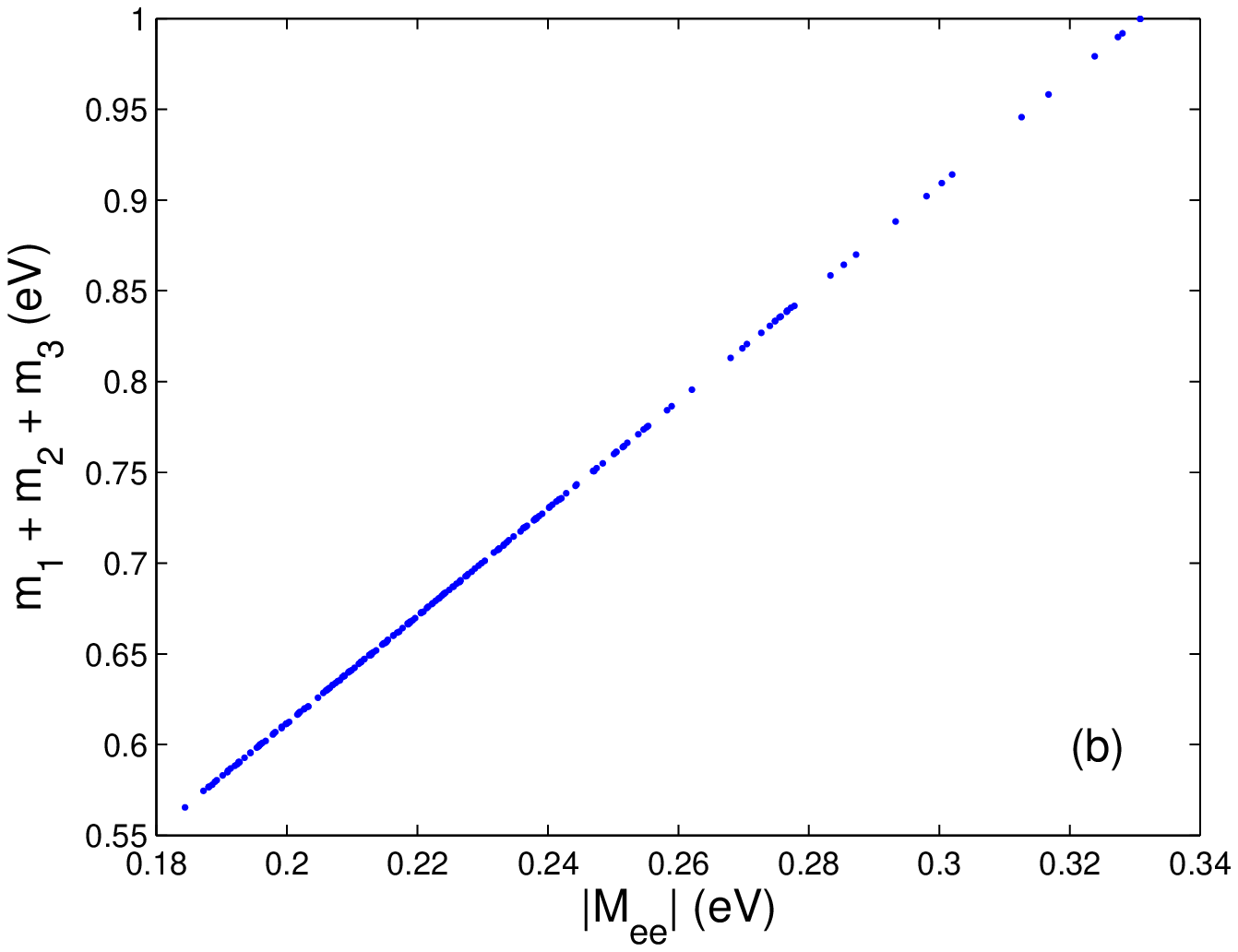} }

\caption{For model 1,
we present (a) $\cos\delta$ \textit{vs.}\ $m_{\beta\beta}$
and (b) the sum of the three neutrino masses
\textit{vs.}\
$m_{\beta\beta}$.
All points shown were found requiring a $1\sigma$ fit to the data of Fogli 
\textit{et al.}~\cite{fogli}.
Model~1 allows an inverted mass spectrum only.}
\label{fig:mod1}
\end{figure}

Model~1 is compatible only with an inverted neutrino mass spectrum.
The neutrino masses must be quite high, of order 0.2~eV or more for each
neutrino, which risks violating cosmological
bounds~\cite{cosmology}.
The phase $\delta$ is close to $\pi$ ($\mathrm{cos}\,\delta < -0.5$),
which agrees with the
(valid at the $1\sigma$ level)
preference found in ref.~\cite{fogli}. The corresponding scatter plots
are presented as figure~\ref{fig:mod1}.

\begin{figure}
\centerline{
\epsfysize=6cm \epsfbox{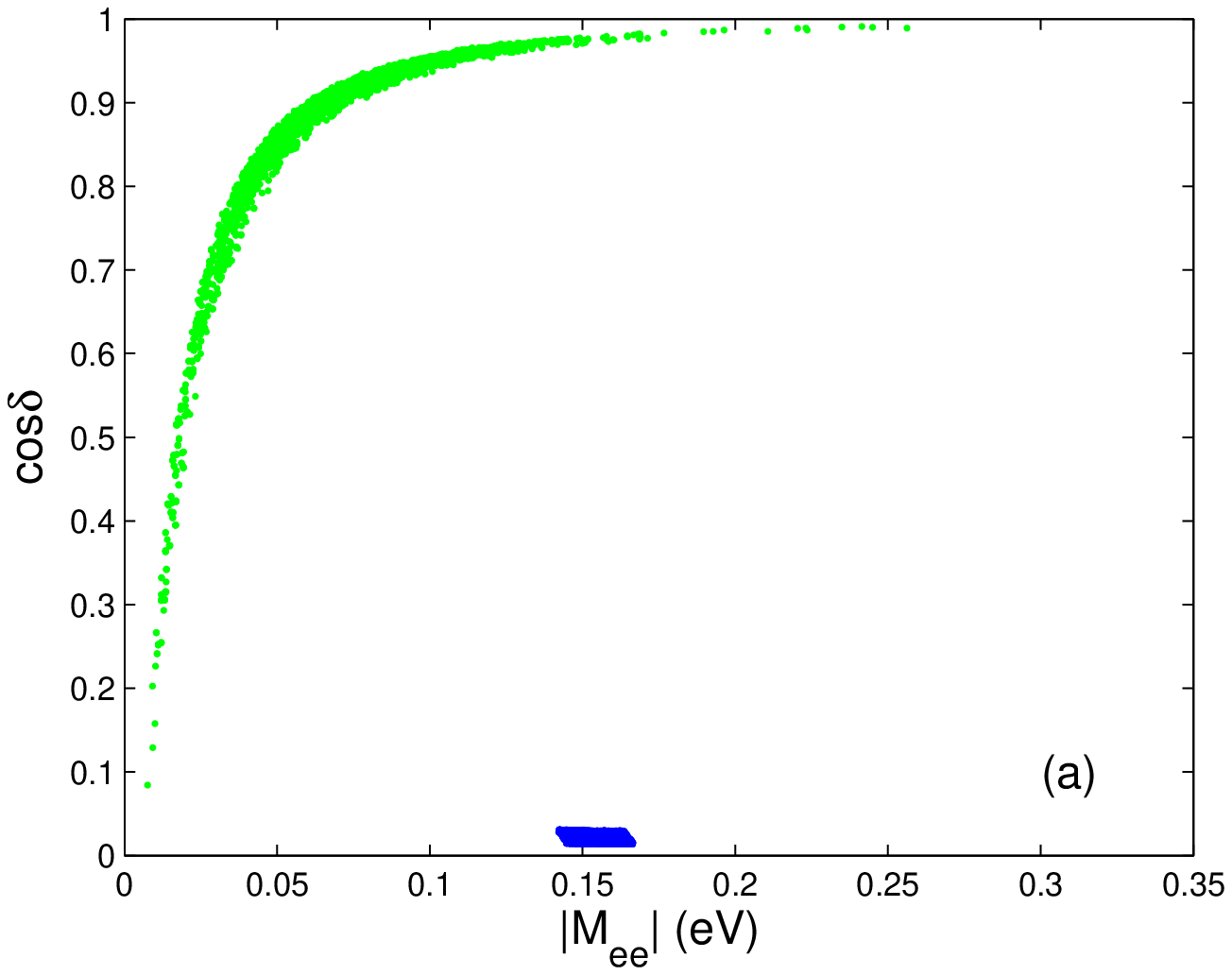}
\epsfysize=6cm \epsfbox{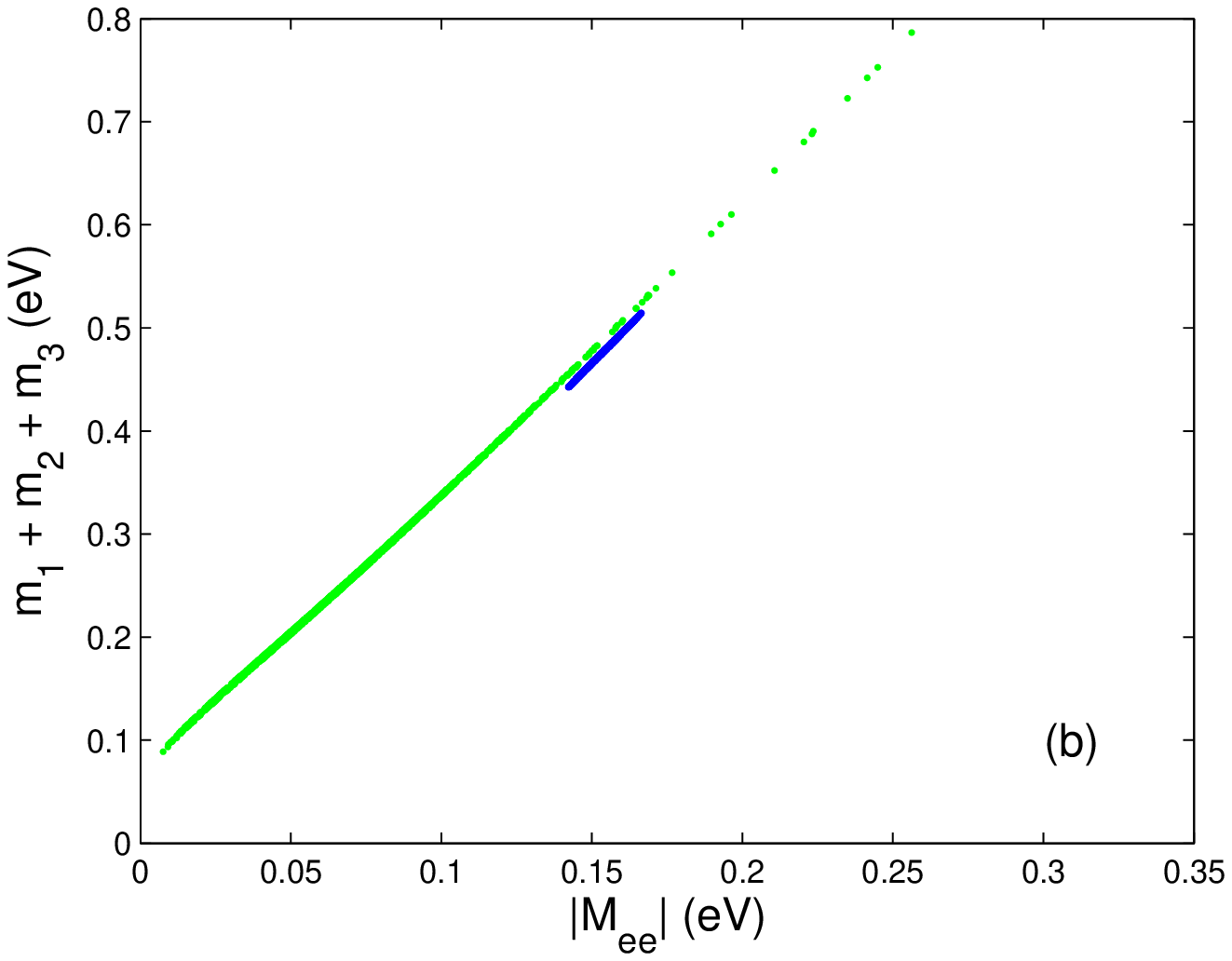}
}
\caption{
For model 2,
we present (a) $\cos\delta$ \textit{vs.}\ $m_{\beta\beta}$
and (b) the sum of the three neutrino masses \textit{vs.}\ $m_{\beta\beta}$.
All points shown were found requiring a $1\sigma$ fit to the data of Fogli
\textit{et al.}~\cite{fogli}.
The green
(light grey)
points correspond to a normal mass spectrum,
while points corresponding
to an inverted mass spectrum are shown in blue
(black).
}
\label{fig:mod2}
\end{figure}

For model~2 both a normal and an inverted neutrino mass spectrum
are possible.
However,
if the spectrum is inverted then
$\delta$ must be very close to $\pi/2$ or $3\pi/2$
and the three neutrinos must be almost degenerate
with masses around 0.15 eV,
which is disfavoured
by some of the recent cosmological bounds~\cite{cosmology}.
For a normal spectrum neutrino masses may be lower
and there is a close correlation between $\cos{\delta}$
and the scale of neutrino masses.
Notice that for model~2 $\cos{\delta}$
is always positive, which disagrees with the findings of
ref.~\cite{fogli} at the $1\sigma$ level;
at the $2\sigma$ level, though, $\cos{\delta}$ is
already unconstrained in ref.~\cite{fogli}, so this should not be
considered a severe handicap of our model.
Plots for this case are given in figure~\ref{fig:mod2}.
\begin{figure}
\centerline{
\epsfysize=6cm \epsfbox{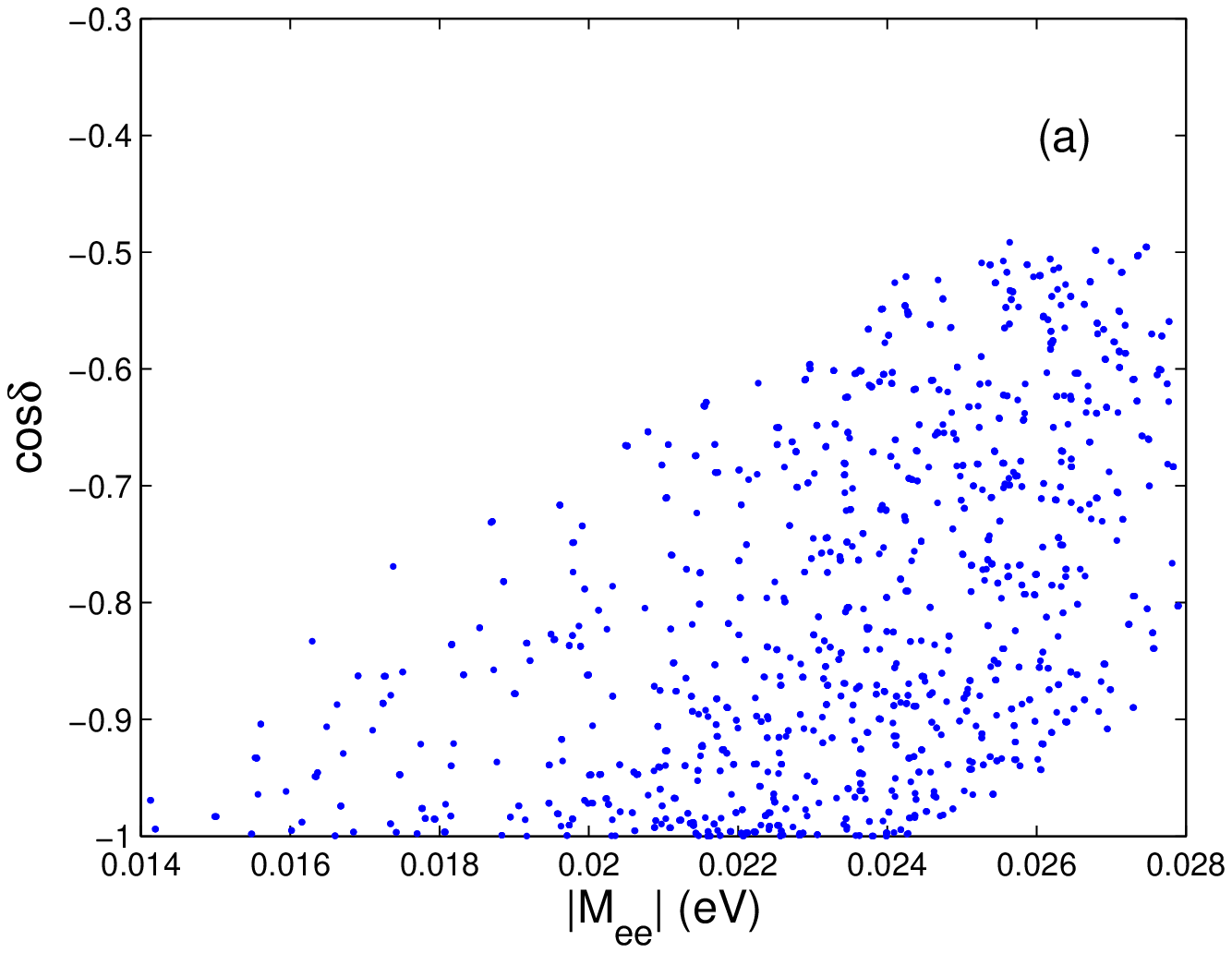}
\epsfysize=6cm \epsfbox{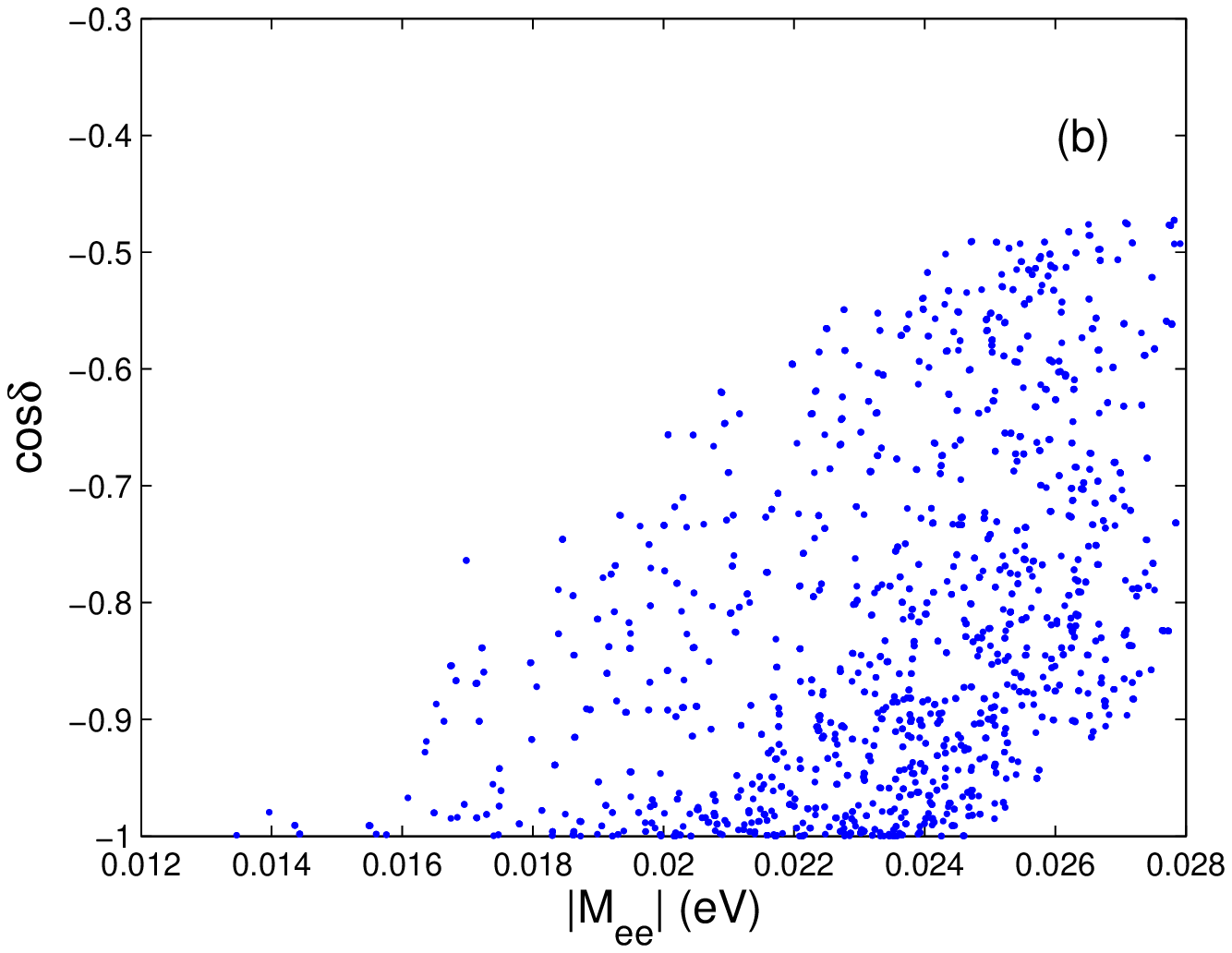}
}
\centerline{
\epsfysize=6cm \epsfbox{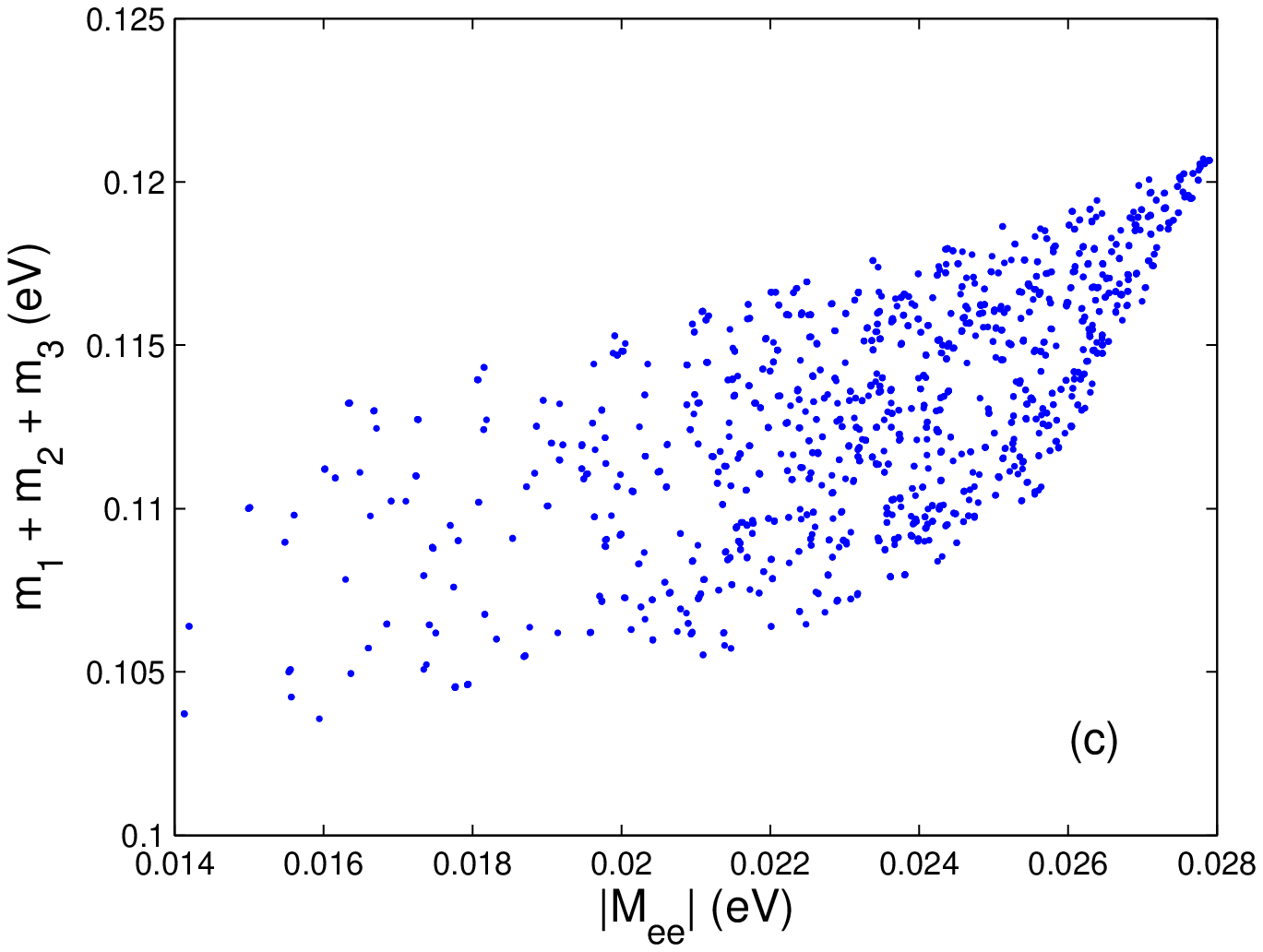}
\epsfysize=6cm \epsfbox{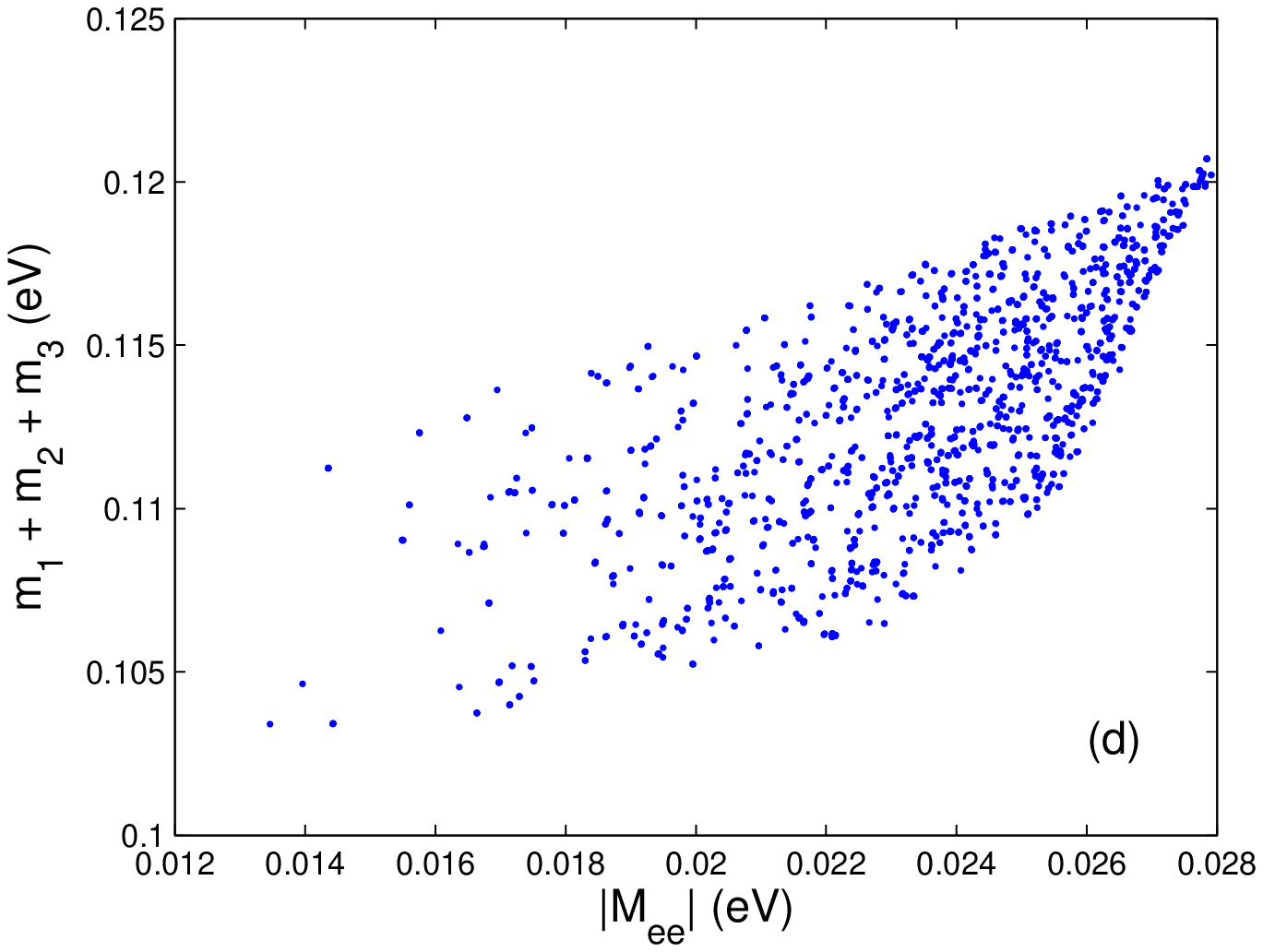}
}
\caption{
Plots for $\cos\delta$ \textit{vs.}\ $m_{\beta\beta}$
in model 3 (plot a) and in model 4 (plot b);
and for the sum of the neutrino masses \textit{vs.}\ $m_{\beta\beta}$
in model 3 (plot c) and in model 4 (plot d).
All points shown were found requiring a $1\sigma$ fit to the data of Fogli
\textit{et al.}~\cite{fogli}.
Models~3 and~4 allow an inverted neutrino mass spectrum only.
}
\label{fig:mod4}
\end{figure}
Models~3 and~4 are,
in practice,
extremely similar.
Only an inverted neutrino mass spectrum is possible for them.
The neutrino masses must moreover lie in a very narrow range:
the lowest mass,
$m_3$,
must lie in between 0.008 eV and 0.012 eV.
This is compatible with all current cosmological bounds.
Moreover,
those models force
$\cos{\delta}<-0.4$,
which is in agreement with the $1\sigma$ range obtained in
ref.~\cite{fogli}---see
eqs.~(\ref{deltarange}).
Plots for
models~3 and~4
are presented in fig.~\ref{fig:mod4}.

\begin{figure}
\centerline{
\epsfysize=6cm \epsfbox{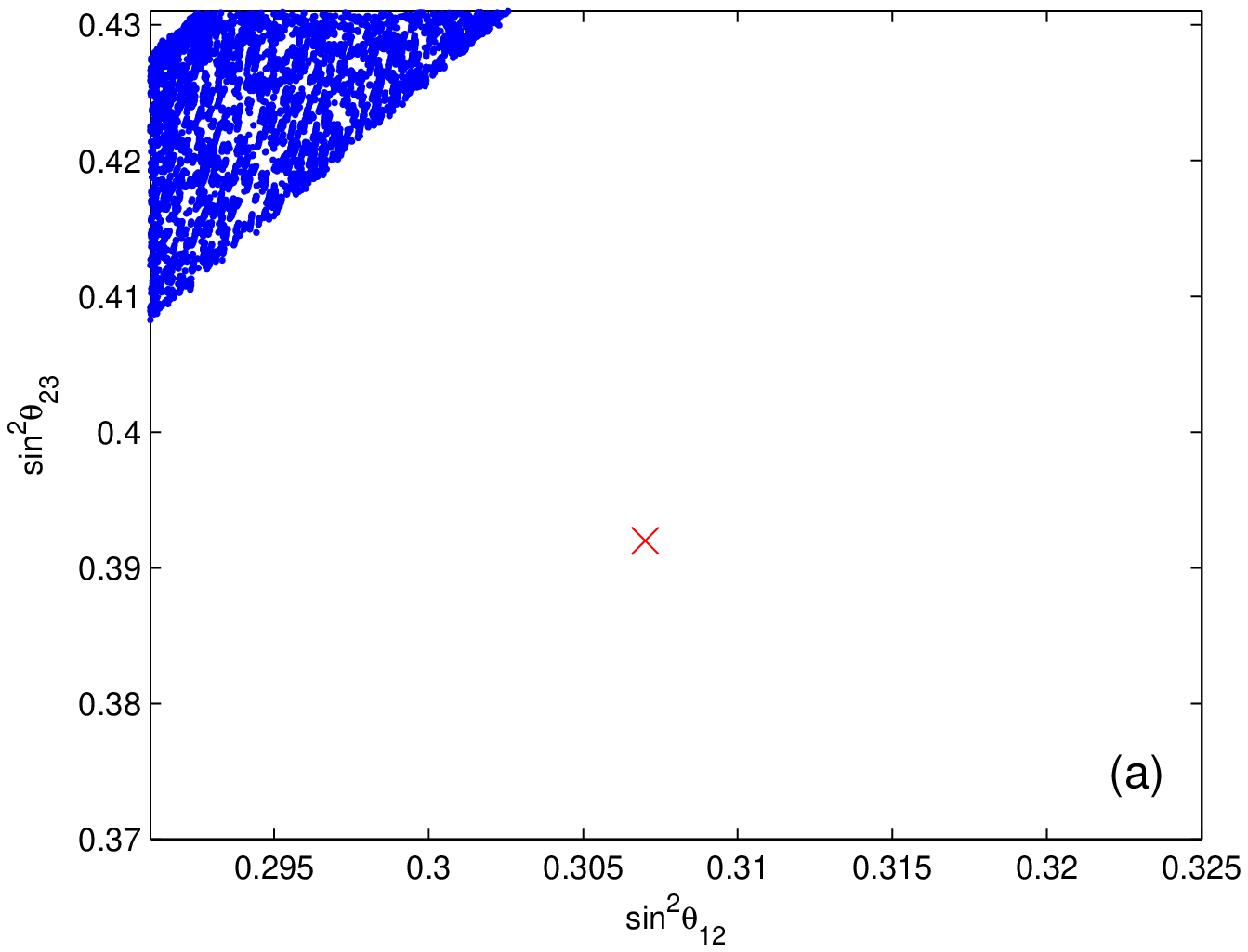}
\epsfysize=6cm \epsfbox{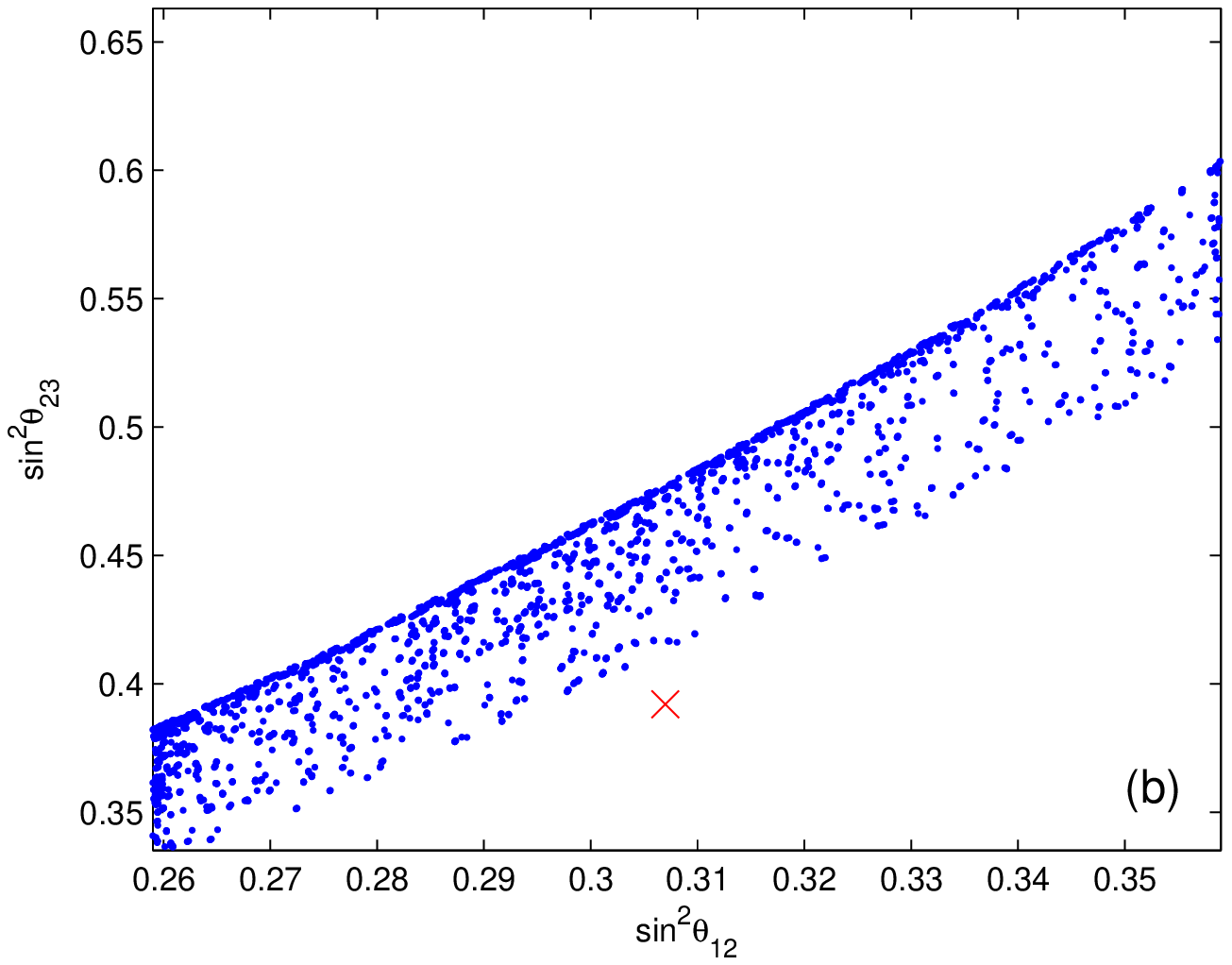}
}
\caption{
The correlation between $\sin^2{\theta_{23}}$
and $\sin^2{\theta_{12}}$ in model~5.
In plot (a) all the physical quantities were forced to lie
inside their $1 \sigma$ ranges of ref.~\cite{fogli}
while in plot (b) the $3 \sigma$ ranges were used instead.
The cross in both plots marks the best-fit point of ref.~\cite{fogli}.
Model~5 only allows an inverted mass spectrum,
with the Majorana phases and $\delta = \pi$ fixed
(see text).
}
\label{fig:mod5}
\end{figure}

In model~5 the neutrino mass matrix $M$ is effectively real,
apart from the unphysical phases of $a$,
$b$,
and $c$;
neglecting those phases,
$M$ may be diagonalised by a real orthogonal matrix $O$ as
$O^T M O = \diag \left( m_1, - m_2, m_3 \right)$.\footnote{We have found that
only this choice of Majorana phases works:
the largest eigenvalue of $M$ in absolute value,
\textit{viz.}\ $m_2$,
must have sign opposite to the one of the other two eigenvalues.}
The smallest neutrino mass is $m_3$,
since only an inverted spectrum works in model~5.
In that model the neutrino masses are exceedingly constrained:
the sum of the light-neutrino masses must be
$\left( 0.110 \pm 0.003 \right)$~\text{eV}
and $m_{\beta \beta} = \left( 0.020 \pm 0.001 \right)$~\text{eV}.
The matrix $O$ is
characterized by Dirac phase $\delta = \pi$.
Besides these predictions,
model~5 predicts a correlation
between the mixing angles $\theta_{12}$ and $\theta_{23}$,
which we depict in
the scatter plots in fig.~\ref{fig:mod5}.
One sees in fig.~\ref{fig:mod5}~(a) that model~5 is still viable,
although rather marginally,
at the $1 \sigma$ level.

\section{Summary}
\label{sec:conclusions}

In this paper we have constructed
five flavour models for the lepton sector.
They are two-Higgs-doublet models
with a standard $\Z_2$ symmetry
which changes the sign of one of the doublets,
or else with a $CP$ symmetry under which one of the doublets is odd.
The models are not necessarily supersymmetric
and use no soft breaking of symmetries,
just spontaneous breaking.
Since they are seesaw models,
they include an extra high scale at which
we introduce
two real scalar gauge singlets
and three right-handed neutrinos.
Models 1--4 have five physical parameters each,
while model~5 only has four parameters.
They all fit very well the phenomenological data
for the lepton mixing angles
and for the neutrino mass-squared differences.
With the exception of model~2,
all our models predict an inverted neutrino mass spectrum
and a negative $\cos{\delta}$.
In models~1 and~2 the neutrino mass scale tends to be quite high,
in possible conflict with cosmological bounds,
but in models~3, 4, and~5 the neutrinos are fairly light.
Model~5 furthermore predicts
a well-defined correlation between the mixing angles
$\theta_{12}$ and $\theta_{23}$.

\vspace*{5mm}

\section*{Acknowledgements:}
We thank Walter Grimus for reading and commenting
on a preliminary version of the manuscript.
The work of PMF is supported in part by the Portuguese
Foundation for Science and Technology (FCT)
under contract PTDC/FIS/117951/2010,
by FP7 Reintegration Grant number PERG08-GA-2010-277025,
and by PEst-OE/FIS/UI0618/2011.
LL
is supported by Portuguese national funds through FCT
project PEst-OE/FIS/UI0777/2011,
and also through the projects PTDC/FIS/098188/2008,
CERN/FP/123580/ 2011,
and the Marie Curie Initial Training Network ``UNILHC'' PITN-GA-2009-237920.
The work of POL is supported by the Austrian Science Fund (FWF)
through the project P~24161-N16.

\end{document}